\documentclass[conference]{IEEEtran}
\usepackage{cite}
\usepackage{verbatim}
\usepackage{amsmath,amssymb,amsfonts}
\usepackage{algorithmic}
\usepackage{graphicx}
\usepackage{textcomp}
\usepackage{xcolor}
\usepackage{dblfloatfix}    % To enable figures at the bottom of page
\def\BibTeX{{\rm B\kern-.05em{\sc i\kern-.025em b}\kern-.08em
    T\kern-.1667em\lower.7ex\hbox{E}\kern-.125emX}}
\begin{document}

\title{Technical Debt in Data-Intensive Software Systems\\
}

\author{\IEEEauthorblockN{Harald Foidl}
\IEEEauthorblockA{\textit{Department of Computer Science} \\
\textit{University of Innsbruck}\\
Innsbruck, Austria \\
harald.foidl@student.uibk.ac.at}
\and
\IEEEauthorblockN{Michael Felderer}
\IEEEauthorblockA{\textit{Department of Computer Science} \\
\textit{University of Innsbruck}\\
Innsbruck, Austria \\
michael.felderer@uibk.ac.at}
\and
\IEEEauthorblockN{Stefan Biffl}
\IEEEauthorblockA{\textit{Institute of Information Systems Engineering} \\
\textit{Technical University of Vienna}\\
Vienna, Austria \\
stefan.biffl@tuwien.ac.at}

}

\maketitle

\begin{abstract}
The ever-increasing amount, variety as well as generation and processing speed of today's data pose a variety of new challenges for developing Data-Intensive Software Systems (DISS). As with developing other kinds of software systems, developing DISS is often done under severe pressure and strict schedules. Thus, developers of DISS often have to make technical compromises to meet business concerns. This position paper proposes a conceptual model that outlines where Technical Debt (TD) can emerge and proliferate within such data-centric systems by separating a DISS into three parts (Software Systems, Data Storage Systems and Data). Further, the paper illustrates the proliferation of Database Schema Smells as TD items within a relational database-centric software system based on two examples. 
\end{abstract}

\begin{IEEEkeywords}
Technical Debt, Data-Intensive Software System, Database Smell, Data Engineering
\end{IEEEkeywords}

\section{Introduction}\label{Intro}
The rapid increase of data storage capacity combined with recent advancements in information technology (e.g. Internet of Things (IoT), Cyber-Phyiscal Systems, Cloud Computing) is causing the amount of today's data to increase at an unprecedented speed. 

Consequently, the design and development of DISS gained increasing attraction and has become an important sub-discipline of software engineering (SE) \cite{Hummel.etal2018}. Loosely based on  \cite{Mattmann.etal2011,Chen.Zhang2014} and \cite{Kleppmann2017} we describe a DISS as a system that especially  
\begin{itemize}
\item processes (e.g. data manipulation and transformation), 
\item writes (e.g. data generation and redistribution) \textit{and}
\item analyzes (e.g. exploratory data analysis) data \textit{as well as} 
\item learns (e.g. through apply learning algorithms) from data
\end{itemize}
in addition to rather general aspects as collecting, storing and visualizing data.
 
These data are typically complex and heterogeneous (i.e. different data types, structures and sources) and of large volume. Furthermore, these data characteristics pose a variety of new challenges (e.g. multidisciplinary teams, weak tool support) for developing DISS that affect the whole software development life cycle (i.e. analysis, design, implementation, testing and maintenance phase) \cite{Anderson2015,Hill.etal2016}. 

The development of DISS typically requires the integration of a multitude of frameworks where each is specialized to perform a specific set of tasks in the system. Hence, DISS typically consist of heterogeneous software architectures where \textbf{software systems} (e.g. traditional software and machine learning applications) and \textbf{data storage systems} (e.g. relational or NoSQL databases and distributed filesystems) have to be integrated in order to interact seamlessly with each other. As those software and data storage systems are mainly fueled by data, \textbf{data} can be seen as a further core element of DISS. Hence, the rest of this paper considers a DISS encompassing a software systems, data storage systems as well as a data part.  

Nevertheless, developing DISS is often done under severe pressure and strict schedules based on today's demand for steadily decreasing time to market. Thus, developers often have to make technical compromises to meet business concerns \cite{Lim.etal2012}. These trade-offs (e.g. system quality vs. development speed) are typically referred as TD.  TD can be described as design or implementation constructs that are beneficial in the short-term (e.g. shortened time to market, high development speed) but incur a debt in form of a technical context which may lead to significant problems in the long-term of a software system (e.g. cost overruns, inability to add new features, project cancellations) \cite{Li.etal2015,Avgeriou.etal2016,Rios.etal2018}. 

However, most research on TD focused on SE architecture, design and artifacts (e.g. source code) \cite{Li.etal2015,Rios.etal2018}. To the best of our knowledge, TD in the context of DISS was not explored in much depth until now. Given the continuous increasing demand for utilizing data in science and industry, we claim that investigating TD in this context becomes an essential need. 
 
Due to its heterogeneous nature, we claim that TD can be incurred in different parts (i.e. software systems, data storage systems, data) of a DISS. Furthermore, we argue that the interplay and entanglement between data, diverse data storage and software systems within a DISS enables TD introduced in one part to further affect other parts. Thus, shortcuts taken in one part can lead to long-term problems in other parts of a DISS. Hence, consequences of shortcuts may have to be paid by experts from other disciplines due to the multidisciplinary nature \cite{Anderson2015,Hummel.etal2018} necessary for developing and operating such data-centric systems (e.g. database design shortcut affects software engineers).

\begin{figure*}[!b]
\centering
\includegraphics[width=6.6 in]{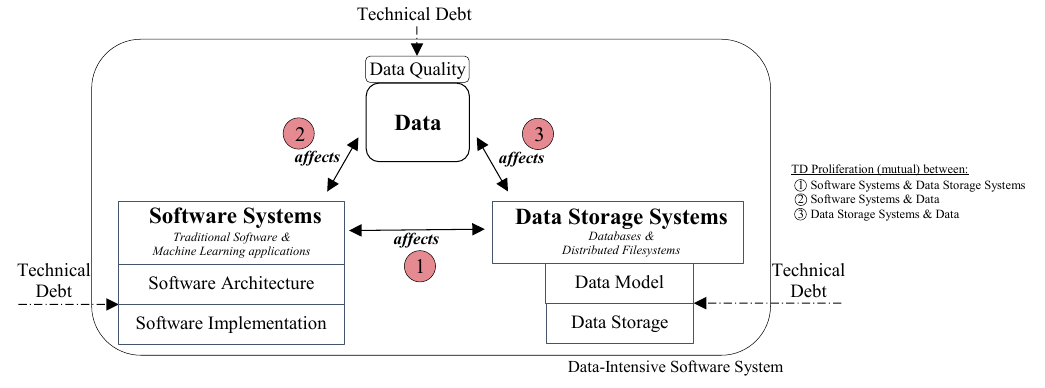}
\caption{Technical Debt Emergence \& Proliferation within Data-Intensive Software Systems}
\label{fig1}
\end{figure*}

In this position paper, we propose a conceptual model that illustrates where TD can arise in DISS. Further, the model shows how incurred debt may proliferate to other parts of the system. The remainder of this paper is structured as follows. Section \ref{Rel.Work} provides a brief overview of related work of TD in the context of data-centric systems. Section \ref{TD-Model} presents a conceptual model which illustrates where TD can emerge in a DISS and further outlines its possible proliferation to the rest of the system. Afterwards, Section \ref{TD-Examples} illustrates the proliferation of two database schema smells as exemplary TD items within a simple DISS. Finally, Section \ref{Outlook} presents possible future work and Section \ref{Conclusion} concludes the paper.

\section{Related Work}\label{Rel.Work}
A considerable amount of literature has been published on TD. The current state of research on TD can be looked up in a tertiary study by \cite{Rios.etal2018} as well as in systematic mapping studies by \cite{Li.etal2015} and \cite{Alves.etal2016}.

There are relatively few contributions in the area of TD related to data-centric environments. Some work was published on TD in data-driven machine learning systems \cite{Sculley.etal2015,Breck.etal2017}. Moreover, TD carried by databases attracted the attention of some researchers (e.g. \cite{Weber.etal2014,Albarak.Bahsoon2016}). Collectively, these contributions focused mainly on TD related to one part of DISS. Hence, one may suppose to provide a holistic view about TD in DISS. Therefore, the next section will present a conceptual model that illustrates the emergence and proliferation of TD within DISS.

\section{Technical Debt Emergence \& Proliferation within Data-Intensive Software Systems \\ - A conceptual Model}\label{TD-Model}
Consequences of TD can be described in terms of cost, value, schedule or quality impact \cite{Avgeriou.etal2016}. Cost, value and schedule can be impacted either positively or negatively depending on the considered time period. For example, schedule can be positively impacted by a workaround in the short term but negatively in the long term when maintenance tasks are delayed due to increased complexity introduced by the workaround. In contrast, incurred debt always decreases the quality of a system in form of structural quality issues regardless of the time period considered \cite{Tom.etal2013}. Therefore, we start discussing the emergence and proliferation of TD within DISS from a quality engineering perspective in this section. More details on the cost impact of TD in DISS will be given in Section \ref{Outlook}. Figure \ref{fig1} shows a model that graphically illustrates the following discussion of the emergence and proliferation of TD within DISS.

Our argumentation is based on the idea to separate a DISS into three parts where TD can mainly emerge and impact the other parts. We separate a DISS based on the idea of Separation of Concerns into its two main technical components (rectangles in Figure \ref{fig1}), named \textbf{Software Systems} and \textbf{Data Storage Systems}. Additionally, as \textbf{Data} are a further core element in DISS, we suggest to define them as third part (rounded rectangle in Figure \ref{fig1}). Within a DISS, software systems typically encompass different processing and computing engines (e.g. Apache Spark, Cascading), machine learning frameworks (e.g. Apache Mahout) and programming models (e.g. MapReduce, Dryad) and languages (e.g. Java, Python). Hence, software systems compromise \textit{Traditional Software} applications as well as intelligent \textit{Machine Learning} applications. Data storage systems comprise different \textit{Databases} (e.g. column-oriented (e.g. Cassandra), relational (e.g. MySQL), document (e.g. MongoDB)) and \textit{Distributed Filesystems} (e.g. HDFS). 

Despite its common usage, TD is defined differently among researchers and comprises different debt types (e.g. architecture, documentation, test \cite{Li.etal2015}). Our understanding of TD in this paper is close to that of \cite{Avgeriou.etal2016}. According to them, TD in its narrowest form (i.e. not considering its consequences) is a collection of design or implementation constructs. Thus, we argue that debt in software systems within a DISS mainly emerges due to \textit{Software Architecture} (e.g. Design Patterns) or \textit{Software Implementation} (e.g. Configuration) decisions. Further, TD accrued in the data storage systems part of DISS is mainly based on \textit{Data Model} (e.g. Data Schema) or \textit{Data Storage} (e.g. Indexing) constructs. Considering the data part, we assert that TD can emerge due to data quality issues. Although data quality is either an implementation nor design construct, we state that bad \textit{Data Quality} is a form of TD that enters a DISS through the data part.

Going into debt at one part of systems consequently can lead to consequences within this part. However, we argue that incurred debt at one part of the system can also affect the remaining parts of the system (illustrated by the numbers 1, 2 and 3 in Figure \ref{fig1}). Accrued debt in the software systems part may affect the data in terms of decreasing quality \textbf{(2)}. For example, skipped constraint checking in the software implementation can lead to bad data quality. Further, inappropriate software architecture constructs can lead to negative consequences within the data storage systems \textbf{(1)}. Misusing a software programming model because there is no time to get used to a model that better fit to the requirements may lead to a decrease of the performance of a data storage system. In the same way, inappropriate design decisions within data storage systems may affect the software systems in a negative way \textbf{(1)} (e.g. complex data processing and querying necessary) or lead to a decrease of data quality \textbf{(3)} (e.g. redundant data). According to \cite{Theodoropoulos.etal2011}, we further argue that data can carry TD in form of bad data quality. As data is moving through the whole system, it can generate new TD in other parts. Bad data quality may foster workarounds in data movement logic or requires intensive data curation which increases the complexity and maintainability of the whole system \textbf{(2)(3)}. Further it can put the analytic integrity of the software systems at risk because the behavior of intelligent algorithms is not specified in code as in traditional software applications but is learned from the data \textbf{(2)}.

\section{Technical Debt Proliferation within a simple Data-intensive Software System \\ - Examples based on Database Schema Smells}\label{TD-Examples}

This section presents two examples that illustrate the model presented above. In detail, the examples aim to show how TD can proliferate within a simple DISS. For sake of comprehensibility, a relational database-centric software system (as an exemplary main part of a simple DISS) is used to demonstrate the proliferation of TD. 

Smells are according to \cite{Alves.etal2016} the most analyzed TD indicators in SE. Further, their validity as instances of TD in software systems was empirically proofed \cite{Zazworka.etal2011}. Hence, we use two of the recently proposed database schema smells by \cite{Sharma.etal2018} as exemplary TD items in the data storage systems part for illustrating their proliferation within a relational database-centric software system. According to \cite{Sharma.etal2018}, database schema smells occur due to the violation of recommended best practices and poor schema design. For demonstrating the TD proliferation as described in the model above, the corresponding number (1, 2 or 3) is annotated in parenthesis after the first sentence in the related paragraph of the examples. As not stated differently, the examples are based on \cite{Karwin2010,Albarak.Bahsoon2016,Sharma.etal2018}.

\subsection{Missing constraints (MC)} This smell describes the situation where referential integrity constraints (RIC) are not declared in a database schema. Reasons such as making the database design simple or flexible, avoiding data update and deletion conflicts or missing implementation knowledge are causes why such constraints are not declared.

The absence of such constraints can lead to serious data quality issues (3). Data values can be entered in child tables that are not referenced to values in the parent table which leads to inconsistent data. Such records that have no corresponding parent rows (orphaned records) may get lost because they never get returned in queries. Further, records in parent tables may have no related records in child tables due to the missing relation.  

However, to avoid these data quality issues software developers are often forced to annotate RIC in the application code (1). This code typically locks tables before updating data or contains several additional queries for checking the existence of referenced values. Further, as typically many different software applications are interacting with the database it can happen that not all annotate RIC in an appropriate way. Hence, additional quality checks have to be implemented to find and correct orphaned records. 

\subsection{Metadata as data (MD)} This smell arises when the EAV (Entity-Attribute-Value) pattern is used to store metadata (attributes) as data. By applying this smell, there is no need to add further columns and hence to change the schema when new attributes have to be stored. Thus, developers often apply this smell as a form of shortcut to gain flexibility.

Nonetheless, the application of this smell may create severe data quality problems (3). Due to the fact that all possible attributes have to be stored in the attribute column, no constraints (e.g. DATE data type, maximum length or RIC) can be declared for this column. Hence, invalid data and attributes that represent the same information but are named differently are not rejected which leads to serious data integrity problems. 

The drawbacks of this smell also affect the software systems (1). Often, software developers have to pay off the problems mentioned above by ensuring data quality checks in the application logic. Further, if mandatory attributes (e.g. no null values) are necessary, software developers have to ensure this in the application code. Additionally, querying an entity with all its attributes requires a rather complex query with a lot of joins. 

These two examples showed that already rather simple shortcuts taken by developing the database can affect the software system and further impact the data in a negative way. This was also explicitly noted by \cite{Albarak.Bahsoon2016} who state that database debt "can ripple to the data it holds and to the applications on top of the database". In the next section, possible ideas on estimating the cost impact of TD in DISS and future research directions are proposed.

\section{Outlook and Research Agenda}\label{Outlook}
Typically, TD is measured in terms of cost impact based on three metrics principal, interest and interest probability \cite{Seaman.Guo2011}. According to \cite{Seaman.Guo2011}, the principal on a debt describes the effort required to eliminate this debt (e.g. refactoring). Interest refers to the penalty that may have to be paid in the future due to the presence of a debt (e.g. increased complexity and maintenance effort) \cite{Seaman.Guo2011,Li.etal2015}. The probability that this penalty has to be paid is described by interest probability \cite{Seaman.Guo2011}. However, considerably more work will need to be done to apply these metrics for managing debt within DISS. As demonstrated above, TD within DISS should not be considered in isolation. Incurred debt at one part of the system is likely to affect another parts and hence creates interests on them. These interests should be considered before shortcuts are taken and debt is incurred. Further, when accumulated debt should be paid off it is necessary to estimate the principal on this debt by considering all parts of a DISS. Taking the examples illustrated in this paper, in addition to refactoring the database also the adaption of the software system and the necessary migration of the data should be considered.
 
Future studies should concentrate on the proliferation of TD within DISS based on the interaction of its three parts. Hence, appropriate techniques, methods, frameworks and tools are needed to support the estimation of interest and principal in a DISS. In the future, we plan to validate our proposed model and to conduct empirical studies to determine interests of incurred debt in DISS. 

Additionally, as relational databases play a vital role in most of today's information systems and are typically used in DISS, we propose that further work should be undertaken to investigate and quantify the cost and quality impact of database schema smells within DISS. As a first step and possible direction for further future work, one could consider to adapt already existing TD assessment frameworks (e.g. \cite{Marinescu2012}) to the domain of database design.

Moreover, the high degree of interplay and entanglement between the three parts of a DISS provides ideal prerequisites to enable the occurrence of dark debt (DD). First mentioned in a workshop in 2017 \cite{Woods2017}, DD is typically found in complex systems and generates anomalies (e.g. complex system failures) that emerge from the unforeseen interactions within a system. Following, DISS would provide a promising context where DD can be investigated.

\section{Conclusion}\label{Conclusion}
This position paper has discussed the emergence and proliferation of TD within DISS. Therefore, a conceptual model focusing on data, data storage and software systems of DISS was presented. The proliferation of TD in form of two database schema smells was demonstrated on a simple DISS encompassing a relational database-centric software system. It is hoped that this paper will raise the awareness of consequences that incurred debt may has within DISS and fosters further research on this topic.

\bibliographystyle{IEEEtran}
\bibliography{IEEEabrv,literature}

\end{document}